 \documentclass[final,5p,times,twocolumn]{elsarticle}

\usepackage{epsfig}

\usepackage{amssymb}

\journal{Physics Letters B}

\begin{document}

\begin{frontmatter}


\title{MANIFESTATION OF ONE- AND TWO-BODY CURRENTS
 IN LONGITUDINAL AND TRANSVERSE RESPONSE FUNCTIONS
OF THE $^{12}$C NUCLEUS AT $q = 300$~MeV/c}

\author{A.Yu. Buki\corref{cor1} }
\ead{abuki@ukr.net}
 \cortext[cor1]{Corresponding author}

\author{I.S. Timchenko}
\address{National Science Center "Kharkov Institute of Physics and Technology",
1,~Akademicheskaya St., Kharkov, 61108, Ukraine}

\begin{abstract}
The experimental values of longitudinal and transverse response
functions of the $^{12}$C nucleus have been obtained at the
3-momentum transfer $q = 300$~MeV/c. The data are compared with the
calculations made with due regard to the dynamics of all the
nucleons constituting the $^{12}$C nucleus, and also, to the
contributions of both the one-body currents only, and their
combination with two-body currents.
\end{abstract}


\end{frontmatter}


\section{}\label{Sec1}
The exact calculation of the longitudinal $R_L(q,\omega)$ and
transverse $R_T(q,\omega)$ response functions of nuclei with full
consideration of the dynamics of all their constituent nucleons is
one of the challenges in quantum many-body physics. So far, the
response function calculations, which are in fairly good agreement
with the experiment, have been performed only for the nuclei with $A
\leq 4$ (e.g., see refs. \cite{{1},{2}}).

In paper \cite{3}, the "first-principles" calculations for the
functions $R_L(q,\omega)$  and $R_T(q,\omega)$  of the $^{12}$C
nucleus were performed on the basis of the AV18+IL7 combination of
two and three-nucleon potentials and accompanying set of two-body
electromagnetic currents. The Green's function  Monte Carlo methods
and maximum-entropy techniques were used in the calculation. In case
of the longitudinal response function, the consideration of
contributions from one-body currents only, or from a combination of
one- and two-body currents, causes an insignificant change in
$R_L(q,\omega)$ only in the vicinity of the threshold. However,
since the two-body currents generate a large excess of strength in
$R_T(q,\omega)$ over the whole $\omega-$spectrum, the comparison
with the experimental data could be a good test of the calculations.

The calculations of response functions in ref. \cite{3} were
compared with the experimental response functions of $^{12}$C,
determined from the world data analysis of J. Jourdan \cite{{4},{5}}
and, for $q =$ 300 MeV/c, from the Saclay data \cite{{6},{7}}. The
data of the mentioned works differ widely. In view of this, it
should be noted that the experimental data on the functions
$R_L(q,\omega)$ and $R_T(q,\omega)$ of the $^{12}$C nucleus were
obtained in Saclay \cite{{6},{7}} at constant momentum transfers $q$
ranging from  200 to 550~MeV/c. In his papers \cite{{4},{5}},
J.~Jourdan has reanalyzed the primary data from refs. \cite{{6},{7}}
and the measured data obtained at SLAC \cite{{8},{9},{10}}, which
were then used for determining the "world" response function values
of the $^{12}$C nucleus. However, not all researchers were content
with the results of the reanalysis \cite{{4},{5}}. For example,
 J.~Morgenstern and Z.-E.~Meziani
have carried out their own reanalysis of the experimental data for a
variety of nuclei, and demonstrated \cite{11} that the results
changed only insignificantly with the combination of the SLAC and
Saclay data.

It follows from the above that for testing the calculations of
ref.~\cite{3}, there is a need to use other experimental data on the
response functions of the $^{12}$C nucleus, which would be
independent of the ones in refs. \cite{{4},{5},{6},{7}}. These data
are derived in the present work and are used for comparison with the
calculations \cite{3}.

\section{}\label{Sec2}
The present experimental response functions were obtained from
processing the spectra of electrons scattered by $^{12}$C nucleus,
which were measured at the NSC KIPT LUE-300 linac at initial
electron energies $E_0$, ranging from 149 to 208~MeV, through the
scattering angle $\theta = 140^\circ$, and at $E_0$ = 200~MeV and
$\theta =  68^\circ$ to $90^\circ$.

Below we give a short description of the measurement and data
processing procedures for obtaining the experimental $R_L(q,\omega)$
and $R_T(q,\omega)$ values (a more detailed information on the topic
can be found, e.g., in ref. \cite{12}).

The electron beam from the accelerator (current being up to 0.2 $\mu
A$) is incident on the target. The scattered electrons are momentum
analyzed by the spectrometer having the solid angle of
$2.89\times10^{-3}$ sr, and the dispersion of 13.7 mm/percent. In
the focal plane of the spectrometer, the electrons are detected by 8
scintillators, each with an energy acceptance of 0.31$\%$. After
that, the electrons come to organic-glass Cherenkov radiators. The
pulses from the photomultipliers of scintillation and Cherenkov
detectors are registered by a coincidence circuit with a time
resolution of 9 nsec.

The spectral measurements of electrons scattering by nuclei involved
the measurement of the contributions that come to the data from the
background processes, viz., the detector registration of radiation
background in the experimental hall (physical background), and also,
of random coincidences as the pulses from scintillation and
Cherenkov detectors arrive simultaneously at the coincidence circuit
(random coincidence background). The electron scattering by the
target is accompanied by photoproduction of $e^+,e^-$-pairs in the
target substance. The electrons of the pairs form one more
background. This background is measured through reversing the
polarity of the spectrometer magnet, and registering the positron
spectrum, which is identical to the electron spectrum from the
$e^+,e^-$-pairs. The measurement of this sort was performed in our
experiment, but no positrons were observed. Perhaps, that was due to
their low yield under those experimental conditions. To check the
conclusion, the positron yield was numerically estimated under the
conditions of the described measurements. The estimations were
performed using the calculation methods from \cite{13}. As à result
was found that the manifestation of electron-positron background in
our measurements was well below the measurement error.

After taking into account the contributions from different
backgrounds, the spectra were corrected for the radiation-ionization
effects by equations of refs.~\cite{{14},{15}}. The measurement data
were normalized with the coefficient $k = F_2^2(q) / F_1^2(q)$,
where $F_1^2(q)$ represents the nuclear ground-state form factor
values obtained in our measurements, and $F_2^2(q)$ stands for the
data taken from work \cite{16}. At that, the 3\% correction (see
ref.~\cite{17}) to the data of \cite{16} was considered.

The experimental values of the longitudinal $R_L(q,\omega)$  and
transverse $R_T(q,\omega)$  response functions of the nucleus are
determined from the analysis of the inclusive electron-nucleus
scattering cross-sections measured at large and small scattering
angles $\theta$. In this case, the equation from ref.~\cite{18} is
used, which connects the response functions with the twice
differential electron scattering cross-section
$\rm{d}\sigma^2/\rm{d}\Omega\rm{d}\omega$, by the relationship
$$
R_\theta(q,\omega)=\frac{\rm{d}^2\sigma}{\rm{d}\Omega\rm{d}\omega}\left(\theta,E_0,\omega\right)/\sigma_M
(\theta,E_0)
$$
$$
=\frac{q^4_\mu}{q^4}R_L(q,\omega)+
\left[\frac{1}{2}\frac{q^2_\mu}{q^2}+\tan^2\frac{\theta}{2}\right]R_T(q,\omega).
\eqno(1)
$$
Here $R_\theta(q,\omega)$ is the angular response function, $E_0$ is
the initial energy of electron scattered through the angle $\theta$
with the transfer of energy $\omega$, the effective 3-momentum $q =
\{ 4E_{eff}[E_{eff} - \omega]$ $\sin^2(\theta/2) + \omega^2
\}^{1/2}$ and 4-momentum $q_\mu = (q^2-\omega^2)^{1/2}$ to the
nucleus studied; $\sigma_M(E_0,\theta) = e^4 \cos^2 ( \theta/2)/
[4E_0^2 \sin^4(\theta /2) ]$  is the Mott cross-section, $e$ is the
electron charge. The term $E_{eff}$ in the definition of the
effective 3-momentum is the effective energy, which is the sum of
the initial energy $E_0$ and the correction $E_C$ that takes into
account the action of the electrostatic field of the nucleus on the
incoming electron. According to \cite{19}, this correction is
written as $E_C$ = 1.33$Ze^{2}<r^{2}>^{-1/2}$, where $Z$ and
$<r^2>$ are, respectively, the charge and r.m.s. radius of the
nucleus.

To obtain the experimental values of the longitudinal and transverse
response functions, it is essential that the set of equations (1)
should be solved for two angular response functions
$R_\theta(q,\omega)$ measured at large and small electron scattering
angles, but at the same $\omega$ and $q$. However, in the plane of
arguments $q$ and $\omega$, the functions $R_\theta(q,\omega)$ can
have only one point in common. Therefore, for obtaining the
$R_L(q,\omega)$  and $R_T(q,\omega)$ values, the set of
cross-sections for electrons scattered by the nucleus is measured in
experiment, from which, after transformation into the function
$R_\theta(q,\omega)$ by means of certain interpolations and
extrapolations with respect to $q$ and $\omega$, the sought-for
values are obtained (for more details, see, e.g., ref.~\cite{20}).

The described processing of the measured data has resulted in the
experimental values of the functions $R_L(q,\omega)$  and
$R_T(q,\omega)$ of the $^{12}$C nucleus at a constant momentum
transfer $q$ = 300 MeV/c. The data are illustrated in Figs. 1a and
1b, divided by the square of the proton charge form factor
$[G_E^p(q_\mu^2)]^2 $ from ref.~\cite{21}.

\begin{figure}\label{fig1}
\includegraphics[width=7cm]{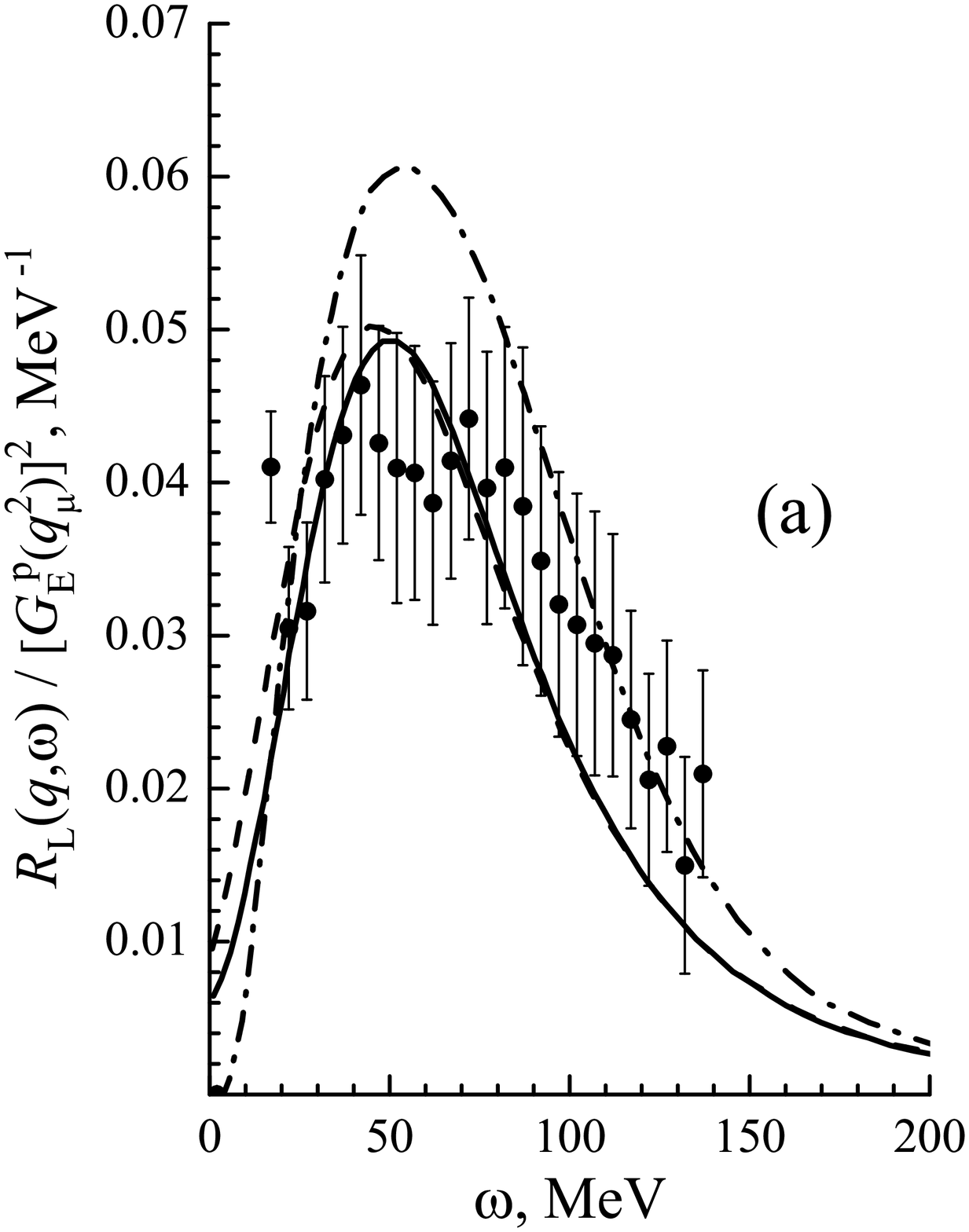}
\includegraphics[width=7cm]{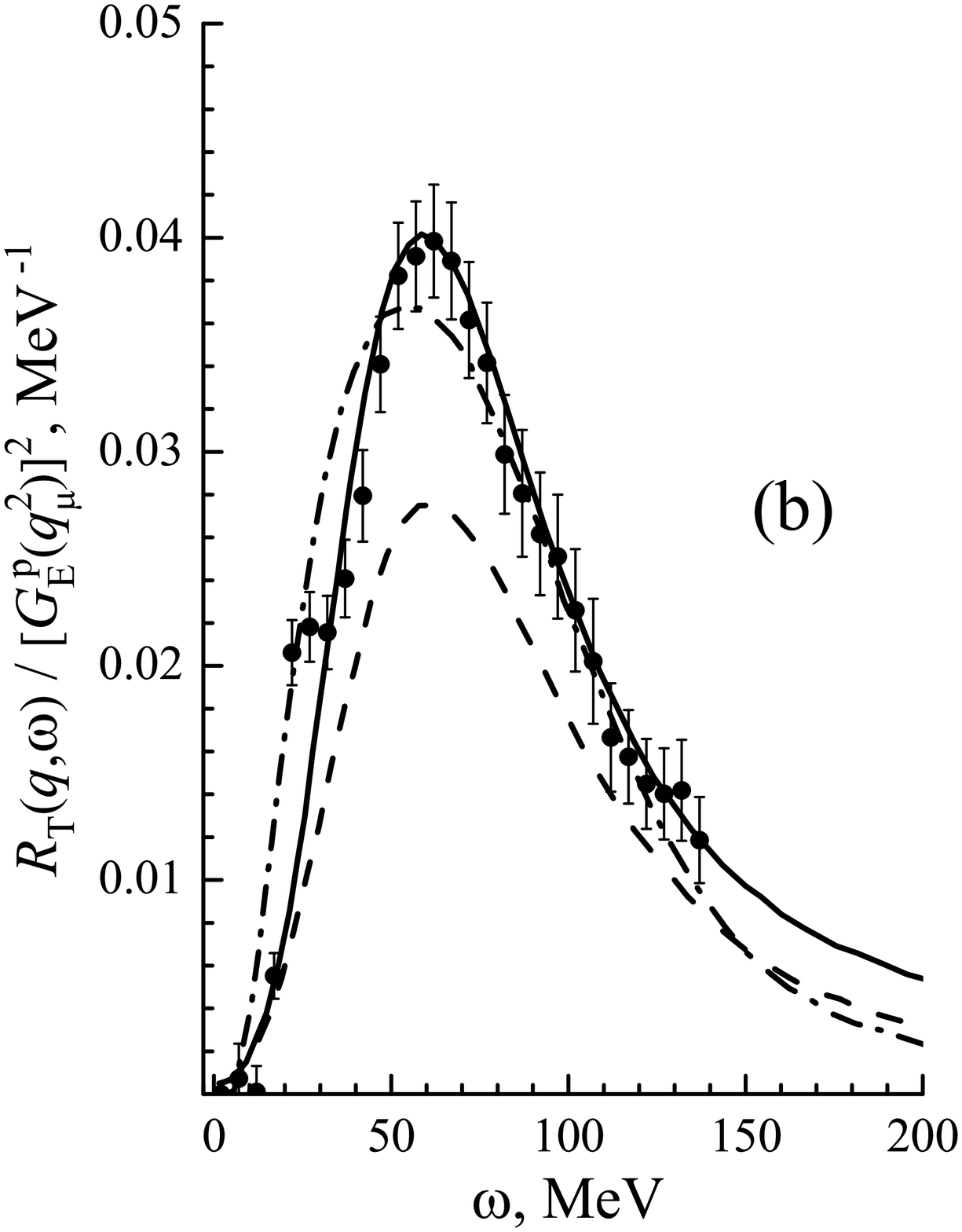}
\caption{$^{12}$C response functions at constant $q = 300$ MeV/c:
(a) longitudinal function $R_L(q,\omega)$; (b) transverse function
$R_T(q,\omega)$.
          The lines show the calculations of work~\cite{3}:
         the dash-and-dot line shows the PWIA calculation;
         the dashed line - the calculation with due regard to one-body currents only;
          the solid line - with due regard to a combination of one- and two-body currents;
         the points show the experimental data of the present work. }
\end{figure}

\section{}\label{Sec3}
Figure 1 shows calculations from work \cite{3} for the longitudinal
and transverse response functions of the $^{12}$C nucleus. The
dash-and-dot line represents the plane-wave impulse-approximation
(PWIA) calculation using the single-nucleon momentum distribution
\cite{22}. The other calculations are based on the realistic dynamic
pattern of the description of nucleus for the cases  with
consideration of only one-body (O1b)  currents in the
electromagnetic operator, and also, with the combination of one- and
two-body currents (O1b-2b). In the last calculations the AV18+IL7
combination of two- and three-nucleon potentials is used.

The comparison of the calculation data for $R_L(q,\omega)$  with the
experimental information points to the fact that the PWIA
calculation overestimates the response value in the longitudinal
component of the quasielastic-scattering peak maximum. In view of
the smallness of the two-body current effect on the longitudinal
response function, the curves for the O1b and O1b-2b calculations
differ only insignificantly. Therefore, none of the calculation
variants can be singled out.

Unlike the $R_L(q,\omega)$ case, in the $R_T(q,\omega)$ case, the
calculations with the contribution of only one-body currents or with
the contribution from combination of one- and two-body currents show
quite a difference, thereby making possible the test of the
calculations. As is seen from Fig. 1b, our data on the function
$R_T(q,\omega)$ are in excellent agreement with the O1b-2b
calculation at all $\omega$ values under study, except in the
near-threshold region, where the contributions of $^{12}$C low-lying
levels were excluded in the O1b and O1b-2b calculations.

Thus, in the present study, we have determined the experimental
functions $R_L(q,\omega)$ and $R_T(q,\omega)$ of the $^{12}$C
nucleus at $q$ = 300 MeV/c. The results are independent of the data
of refs.~\cite{{4},{5}} and \cite{{6},{7}}, which were earlier used
for testing the calculations of ref.~\cite{3}. Our present
experimental values of the response functions under consideration
correspond to the calculation variant of ref.~\cite{3}, in which the
combination of one- and two-body currents was taken into account.

\end{document}